# Quantum phase slip phenomenon in ultra-narrow superconducting nanorings

KONSTANTIN YU. ARUTYUNOV[1,2]*, TERHI T. HONGISTO[1], JANNE S. LEHTINEN[1], LEENA I. LEINO[1], and ALEXANDER L. VASILIEV[3]

[1]NanoScience Center, Department of Physics, University of Jyväskylä, PB 35, 40014, Jyväskylä, Finland

[2] Moscow State University, Institute of Nuclear Physics, Leninskie gory, GSP-1, Moscow 119991, Russia

[3]Kurchatov Institute, 1, Akademika Kurchatova pl., 123182 Moscow, Russia

* konstantin.arutyunov@phys.jyu.fi

The smaller the system, typically - the higher is the impact of fluctuations. In narrow superconducting wires sufficiently close to the critical temperature $T_c$ thermal fluctuations are responsible for the experimentally observable finite resistance. Quite recently it became possible to fabricate sub-10 nm superconducting structures, where the finite resistivity was reported within the whole range of experimentally obtainable temperatures. The observation has been associated with *quantum fluctuations* capable to quench zero resistivity in superconducting nanowires even at temperatures T➔0. Here we demonstrate that in tiny superconducting nanorings the same phenomenon is responsible for suppression of another basic attribute of superconductivity - persistent currents - dramatically affecting their magnitude, the period and the shape of the current-phase relation. The effect is of fundamental importance demonstrating the impact of quantum fluctuations on the ground state of a macroscopically coherent system, and should be taken into consideration in various nanoelectronic applications.

Thermodynamic fluctuations become more important when dimensions of a system are reduced. It has been known for decades that in narrow quasi-1D superconducting channels thermal fluctuations enable zero resistance only at temperatures noticeably below the critical temperature $T_c$ [1,2,3,4]. Quite recently it has been demonstrated that qualitatively different phenomenon - quantum fluctuations - also called *quantum phase slips* – can suppress the dissipationless electric current in ultra-narrow superconducting nanowires with characteristic diameter d~10 nm even at temperatures T$\rightarrow$0 [5,6,7,8,9,10,11,12] setting the ultimate limit for utilization of superconducting nanoelectronic components. Zero resistivity, detected in transport experiments, being probably the most famous 'text-book' manifestation of superconductivity, characterizes the non-equilibrium state of a superconductor. On the contrary, persistent current (PC) circulating in a superconductor exposed to an external magnetic field is the thermodynamically equilibrium characteristic of the system.

The impact of fluctuations on equilibrium properties of a quantum system is a very intriguing topic. Of particular interest is the prediction that in ultra-narrow superconducting nanorings exposed to external magnetic field the quantum fluctuations should induce a gap in the energy spectrum forming a band structure, and degenerate the PC saw-tooth current-phase dependence into a smooth sine-type [13,11] (Fig. 1, dotted lines). Direct measurement of the impact of fluctuations on magnetic momentum of PCs circulating in a nanoscale superconducting sample is a rather demanding experimental task. To our best knowledge, so far the corresponding experiments were reported only sufficiently close to the critical temperature $T_c$ [14,15] where the 'classical' thermal fluctuations provide the dominating contribution [16,17]. The objective of our work was to study the qualitatively different limit: ultra-low temperatures and ultra-narrow samples where the impact of thermal fluctuations can be disregarded, while quantum fluctuations should manifest themselves. For probing PCs we employed solid state tunneling technique (for details see *Methods*). The method provides maximum sensitivity for homogeneous uniform samples, eliminating those ones where the existence of undesired weak link(s) may mimic the effects under interest. The extensive analysis of the whole set of experimental data and the alternative explanations allows us to state, that the most plausible explanation for our findings is indeed related to the quantum phase slip phenomenon. The effect should be taken into consideration in various nanoelectronic applications, and is of fundamental and universal importance reflecting the impact of quantum fluctuations on the ground state of a macroscopically coherent system.

## RESULTS

Let us first outline the main features of the model which supports the experimental findings presented below. The energy levels $E_n$ of a thin-walled superconducting ring threaded by magnetic flux $\Phi$ are given by the set of parabolas (Fig. 1a). If the system follows the ground state, the evolution of the energy in magnetic field is represented in Fig. 1a by the thick solid line with periodicity strictly equal to one flux quantum $\Delta\Phi = \Phi_0 = h/2e$. The scenario is typical for experiments made sufficiently close to the critical temperature $T_c$ [18,14] or to systems containing weak links – SQUIDs. The corresponding dependence of PC on magnetic flux $I_s \sim dE_n/d\Phi$ follows the single flux quantum periodicity (Fig. 1b, thick solid line) with the well-known saw-tooth pattern, which can be smeared if the contribution of thermal fluctuations is essential [14].

At temperatures $T \ll T_c$ the system can be 'frozen' in a metastable state n without relaxing to the neighboring quantum level n±1. In this low temperature limit the periodicity of PC oscillations may significantly exceed one flux quantum [19,20]. Transition to another quantum state, triggered by an 'external disturbance', happens when:

$$G(I_s) \approx \max(k_B T, \mathcal{E}), \qquad (1)$$

where G is the current-dependent energy gap of the superconductor and $\mathcal{E}$ represents the cumulative impact of all non-thermal contributions. In the 'classical' limit $\mathcal{E} \ll k_B T$ Eq.(1) is equivalent to the trivial condition: maximum PC equals the critical current $I_s = I_c(T)$ (Fig. 1b, horizontal line). At sufficiently low temperatures (and negligible external disturbance) one should observe periodicity $\Delta\Phi/\Phi_0 \sim S/\xi \gg 1$ if the loop circumference S is much larger than the superconducting coherence length $\xi$ [20]. PCs approaching sub-critical values broaden the superconducting density of states (DOS) [21] $N(E) = N(0) \mathrm{Re}\{\cos[\theta(E)]\}$ which can be found from the self-consistent solution of the Usadel and gap equations for the pairing potential $\Delta$:

$$E + i\Gamma\cos\theta = i\Delta\cot\theta, \tag{2}$$

$$\Delta = N(0)V_{eff}\int_0^{\Theta_{Deb}}\tanh(E/2k_BT)\,\mathrm{Im}(\sin\theta)\,dE, \tag{3}$$

where $\Gamma$ is the deparing energy; $\theta$ is the pairing angle; $N(0)$, $V_{eff}$ and $\Theta_{Deb}$ are the normal state DOS, pairing strength and Debye energy. If the superconducting loop is a part of a tunnel structure, then the periodic modulation of DOS results in tunnel current oscillations (Fig.1c) [22,23]. Supported by the known expression for $E_n(\Phi)$ [11,13] and the current dependence of the deparing energy $\Gamma(I_s)$ [21], expressions (1)-(3) form a complete set of equations enabling quantitative analysis [22,23]. At magnetic fields much smaller than the critical one the direct deparing by magnetic field $\Gamma(B)$ [21] can be neglected compared to the corresponding current dependence $\Gamma(I_s)$.

For a given material in the 'classical' limit $E \ll k_B T \ll G$ the large period of $I_s$ (or, alternatively, $I_{tun}$) oscillations in units $\Delta\Phi/\Phi_0$ (Fig. 1 dashed lines and Fig. 3a) is determined exclusively by the loop circumference S and does not depend on the cross section $\sigma$ of the wire forming the loop, as long as the condition of one-dimensionality $\sigma^{1/2} < \xi$ is satisfied. In the 'quantum' limit of extremely narrow loops when the rate of quantum fluctuations $E_{QPS} \sim \exp(-\sigma)$ [11] exceeds all other sources of disturbance, the shape of oscillations should demonstrate smooth quasi-sinusoidal pattern [13,11] of much smaller amplitude (Fig. 1, dots and Figs. 3c and 4). Calculation of the tunnel current $I_{tun}(\Phi)$ in presence of quantum fluctuations is not trivial: strictly speaking, Eqs. (2)-(3) are valid only when the pairing potential $\Delta$ is a well-defined parameter, which is not the case in a fluctuation regime. Nevertheless we believe some estimates can be made.

The QPS contribution [5,6,11] becomes important when $k_B T \ll E_{QPS} = \Delta\dfrac{R_Q}{R_N}\left(\dfrac{S}{\xi(\Delta)}\right)^2\exp\left[-S_{QPS}\right] \leq G$

, where $S_{QPS} = A\dfrac{R_Q}{R_N}\dfrac{S}{\xi(\Delta)}$, A - material constant, $R_N$ – normal state resistance of the loop, $R_Q = h/4e^2$ - quantum resistance. Even in the regime of weak quantum fluctuations $E_{QPS} \ll E_R = (\pi^2 h^2 \sigma n_s \ell)/(m^* S \xi_0)$ the dependence $I_s(\Phi)$ already deviates from the saw-tooth [11] (Fig. 3c, right inset):

$$I_s = \left(\frac{2e}{h}\right) E_R \frac{\left(1 - \frac{\pi^2 E_{QPS}^2}{2 E_R^2}\right)}{\pi \sqrt{1 - \left(1 - \frac{\pi^2 E_{QPS}^2}{2 E_R^2}\right)^2 \sin^2\left[\pi \frac{\Phi}{\Phi_0}\right]}} ArcSin\left\{\left(1 - \frac{\pi^2 E_{QPS}^2}{2 E_R^2}\right) \sin\left[\pi \frac{\Phi}{\Phi_0}\right]\right\} \cos\left[\pi \frac{\Phi}{\Phi_0}\right], \quad (4)$$

where e, $\ell$ and m* and $n_s$ are charge, mean free path, effective mass and density of superconducting electrons. To obtain the $I_{tun}(\Phi)$ dependencies to be compared with the experiment Eq. (4) should be solved self-consistently with Eqs. (2)-(3).

Modeling of the 'intermediate' regime $k_B T \ll \mathcal{E} < G$ is, as usual, the most difficult. In this intermediate case, from the very general considerations, one might expect the magnitude and the period of PC oscillations to be reduced, compared to the 'classical' regime, demonstrating variable oscillations due to the stochastic nature of a disturbance quenching the oscillations before the condition $I_s = I_c(T)$ is reached (Fig. 3b). Unless the oscillations are completely suppressed in the limit of a non-negligible 'disturbance' $\mathcal{E} \approx G$, the dependencies $I_s(\Phi)$ and $I_{tun}(\Phi)$ should follow the saw-tooth pattern: linear and curved, respectively (Figs. 1b,c (solid and dashed lines) and Figs. 3a,b). For the utilized dimensions of the loops (Fig. 2) the contribution of the geometrical inductance can be neglected. The energy related to the kinetic inductance $E_L = \Phi_0^2 / (2 L_{kin})$ is always the largest energy scale enabling consideration of the QPS-mediated gap (Fig. 1a, dots) as a small perturbation imposed on the parabolic energy spectrum (Fig. 1a, solid lines). The smallness of the fluctuation rate $E_{QPS}/h$ justifies utilization of Eq. (4) and the employed model which is based on the assumption that the QPS are relatively rare events[5,6].

Summarizing the theoretical background, one may conclude that at low temperatures $T \ll T_c$ for a superconducting loop of a given area and circumference $S \gg \xi$ the reduction of the cross section $\sigma$ of the wire forming the loop should lead to a qualitative changes in the PC oscillation pattern. For 'thick' loops the period $\Delta\Phi/\Phi_0 \gg 1$, the magnitude is large and shape is saw-tooth (Fig. 1, dashed lines) [19,20,22]. While in 'thin' samples the quantum fluctuations should degenerate the PC oscillations into smooth sine-type pattern with much smaller magnitude and strictly one flux periodicity (Fig. 1, dotted lines) [13,11]. For straightforwardness of interpretation the outlined scenario assumes that both the thick and the thin samples are homogeneous metallic loops containing no weak links. However, it should be noted that for high transparency barriers between the superconducting grains the model[13], developed for a loop formed by a chain of weak links, converges to the homogeneous (metallic) loop model [11]. The observation naturally eliminates the distinction between a chain of weak links and a homogeneous superconducting channel in the regime of quantum phase slips.

To test these hypotheses all-superconducting nanostructures with the loop-shaped central electrode (Fig. 2) were fabricated (for details, see *Methods*). In zero magnetic field all structures demonstrated conventional $I_{tun}(V)$ dependencies typical for a S1-I-S2-I-S1 tunnel structure. Text-book fits enable determination of the tunnel resistance, the intrinsic deparing energy $\Gamma_0/\Delta_0 \sim 0.01$ and the effective electron temperature $T_e$ (typically ~70 mK at a base temperature ~50 mK). With the bias voltage V fixed below the gap edge $eV < 2(\Delta_1 + \Delta_2)$ perpendicular magnetic field was slowly swept and the tunnel current was measured $I_{tun}(V=const, B)$.

All 'thick' structures showed pronounced saw-tooth tunnel current oscillations with period $\Delta\Phi/\Phi_0 = const \gg 1$ in a full accordance with the earlier findings [22] and the model [23] (Fig. 3a). The simulated flux dependencies of the tunnel current $I_{tun}$, superconducting order parameter $\Delta$ and gap G, deparing parameter $\Gamma$ and PC $I_s$ are presented in Fig. 3 main panel, left right insets, respectively. As for these 'thick' samples the experimental $I_{tun}(V=const, B)$ dependencies demonstrate constant period, one may conclude that the instrumental-related flux noise is negligible and the spontaneous phase slips can be disregarded. The large period of oscillations corresponds to the condition when the flux-dependent superconducting gap equals the thermal energy $G(\Phi) = k_B T$ (Eq. (1)) (Fig. 3a, left inset). Due to the strong current dependence of the deparing parameter $\Gamma(I_s[\Phi])$[21] at low temperatures $T \ll T_c$ the above condition is quantitatively almost indistinguishable from the complete suppression of superconductivity $\Delta(I_s[\Phi])=0$. The flux dependence of the PC is strictly saw-tooth with the transition to a new quantum state n ('vorticity') very close to the critical value $I_s(\Phi)=I_c$ (Fig. 3a, right inset). Better quantitative agreement between the experimental and simulated $I_{tun}(\Phi)$'s can be achieved taking into consideration the monotonous reduction of the superconducting gap $\Delta(B)$ by magnetic field (noticeable in Fig. 3a) and the solid state heating/cooling phenomenon[24,25]. However, we found the improvement of the fits marginal and being not very important for the main subject of the paper.

All studied structures, including the ones with the thickest loops, can be considered as quasi-one-dimensional: $\sigma^{1/2}<\xi(T)$, with $\xi$ being the superconducting coherence length. The observation enables utilization of the phase slip concept strictly relevant only for a 1D system. In the 'classical' limit $\mathcal{E}<<k_BT$ the reduction of the wire cross section $\sigma$ does not alter the period of oscillations given the structures are sufficiently 'thick'. However, below a certain threshold, $\sigma_q^{1/2}(Al)\sim20$ nm and $\sigma_q^{1/2}(Ti)\sim65$ nm, the magnitude and the period of oscillations rapidly drop demonstrating the saw-tooth pattern with variable period (Fig. 3b). As the flux noise was not providing a noticeable contribution for thicker samples, it is reasonable to assume that the observed variable period for thinner structures is related to some intrinsic size-dependent mechanism of stochastic nature – e.g. thermodynamic fluctuations. Further $\sigma$ reduction degenerates the oscillations into a smooth quasi-sinusoidal pattern with significantly reduced magnitude (Fig. 3c and Fig. 4) and the period $\Delta\Phi/\Phi_0=1$. Note that SEM/AFM analysis did not reveal any breaks or/and pronounced constrictions. The I(V, B=0) dependencies of all samples, including the thinnest ones, did show-up conventional behavior typical for a S1-I-S2-I-S1 tunnel structure. Peculiarities which could be associated with formation of additional junction(s) (e.g. S2'-I-S2'') were detected neither in the original, nor in the ion milled samples. No hysteresis on I(V, B=0) characteristics was observed supporting the absence of overheating and the low-invasiveness of the employed tunneling method: the characteristic tunnel current is below $10^{-10}$ A (Fig. 4), while the corresponding critical current in nanowires with similar diameters is about $10^{-7}$ to $10^{-6}$ A [26].

## DISCUSSION

One can interpret the results assuming that below the threshold cross section $\sigma_q$ either (i) coupling of the system to external disturbance increases; or (ii) sample imperfections (revealed hidden intrinsic and/or generated by sample processing) qualitatively change the properties of the system; or (iii) an intrinsic size phenomenon develops.

First option looks rather improbable as coupling of a nanoloop, isolated by the pair of high-Ohmic tunnel contacts S1-I-S2, should not depend on the wire cross section $\sigma$. Certainly, the employed method of low energetic ion beam milling slightly reduces the area of the probing tunnel junctions[27,28] leading to the increase of the tunnel resistance. However, the effect is small and by no means can explain the hypothetical increase of the coupling of the sample to the noisy external environment below the threshold $\sigma_q$.

Second alternative (sample imperfections) has been given the highest attention. An interpretation based on formation of a weak link – Josephson junction (JJ) – between the strong (massive) superconducting arms of the loop (SQUID-type structure) or a chain of JJs is not credible. First, an extensive SEM/AFM analysis revealed no suspicious sections to be considered as potential weak links with the local critical current density much smaller than in the rest of the loop (Fig. 2). Here we would like to emphasize the importance of titanium data (Fig. 4): the cross sections of the loops are so 'huge' ($\sigma^{1/2} \sim 60$ nm) and the surface roughness, measured by AFM, is so small ($\pm 2$ nm, Fig. 2d, inset) that the existence of a hypothetical JJ is rather unrealistic. The nanostructures were fabricated and processed using the same technique as the nanowires, where an extensive microscopic and elemental analysis confirmed their homogeneity down to significantly smaller cross sections[12]. Second, if occasionally such a hypothetic JJ is formed, the properties of such a system would be unique depending on the particular values of the charging and Josephson energy of that junction. However, the results are well reproduced on samples with close values of the diameter of the wire forming the loop. The observation supports the conjecture that our results deal with a universal size effect, rather than with accidentally formed JJs with random parameters. Third, if at least a pair of S2'-I-S2'' junctions (one, in each arm of the loop) and/or an extra single tunnel junction in the pole region are formed, then the existence of that extra junction(s) should manifest itself as an appearance of a certain peculiarity on the I-V characteristic, which has never been the case. Forth, even if an undetectable hypothetical weak link is formed, the amplitude of the experimentally measured tunnel current oscillations would be infinitely small. The employed method is based on measuring the tunnel current within the locus of the probing tunnel junction. The oscillating in the loop PC, periodically reaching sub-critical values, modulates the DOS leading to the corresponding oscillations of the tunnel current. Thus the method can provide a decent sensitivity only in the case of a relatively homogeneous loop with the uniform current density including the regions of the tunnel probes. Due to the employed fabrication method (multi-angle metal evaporation followed by lift-off lithography) in our samples the regions close to the tunnel probes ('poles') are always slightly wider than the thinner 'equatorial' parts of the loop (Fig. 2), where the probability of formation of a weak link (e.g. constriction) is higher. If the case, within the locus of the tunnel probe the oscillations of the DOS and the corresponding tunnel current would be negligible being insensitive to the modulation of the local current $I_s^{probe}$ by the much smaller critical current of a (remote) weak link $\Delta I_s^{probe} \sim I_c^{weaklink} << I_c^{probe}$. Summarizing, the JJ scenario cannot explain the gradual reduction of the period and the magnitude of the tunnel current oscillations with reduction of the wire cross section (Fig. 3 and 4). As soon as inside the loop a single JJ with much smaller local critical current is formed, the amplitude of the tunnel current oscillations would drop to undetectably small values.

The third option – intrinsic size-dependent disturbance $\varepsilon$ – looks mostly intriguing. Contribution of thermal fluctuations [1,2] can be neglected at temperatures $T << T_c$ [29]. Though the corresponding model [1,2] is not applicable to the ultra-low temperatures of the experiment, one may argue that qualitatively thermal fluctuations should be taken into consideration when the flux-dependent superconducting gap is of the order of the thermal energy $G(\Phi) \sim k_B T$. Due to the very sharp $G(\Phi)$ dependence (Fig. 3a, left inset) the correction might be important only within the very vicinity of the transition to a new quantum state: the very 'tips' of the saw-tooth $I_s(\Phi)$ and $I_{tun}(\Phi)$ oscillations. By no means the thermal fluctuations could explain even qualitatively the changes in the shape, the magnitude and the period of $I_{tun}(\Phi)$ oscillations with the reduction of the loop cross section $\sigma$ (Figs. 3 and 4). Various mesoscopic fluctuations[30, 31] may reduce the local potential barrier quenching formation of the metastable states responsible for the large period oscillations, while applicability of these models to the experiment is not clear.

On the contrary, the recent models considering quantum fluctuations in superconducting nanorings [13,11] seem to be quite relevant. Whether the loop is homogeneous [11], or consists of weakly coupled superconducting grains [13] - both models predict the same dependencies $E_n(\Phi)$ and $I_s(\Phi)$ (Fig. 1, dotted lines). However, as discussed above, in case of weak links [13] the tunnel current oscillations $I_{tun}(\Phi)$ would not be experimentally resolved. We do not have any experimental evidence that the loops in our structures can be considered as 1D chains of JJs with the local critical current significantly lower than in the rest (metallic) part of the loop. The studied samples do not have an ideal (atomically flat) surface, but they are smooth enough to be considered as homogeneous metallic loops with the variation of the diameter below $\pm 3$ nm. Quantum fluctuation scenario is particularly intriguing as the $\sigma_q^{1/2}$ threshold values correspond to the diameters where the broadening of R(T) dependencies has been associated with quantum phase slip effect in aluminium [9,10] and titanium[12] nanowires. Simulations based on QPS model provide reasonable quantitative agreement with experiment (Figs. 3c and 4). Dependence (4) inevitably results in doubling of the tunnel current oscillation period (Figs. 3c and 4, solid curves). The fine structure was not clearly resolved in experiment: in the QPS limit the $I_{tun}(\Phi)$ oscillation magnitude exponentially drops with decrease of the wire cross section $\sigma$ [11] burying the oscillation pattern below the experimental noise $\sim 10^{-13}$ A. Nevertheless the very observation of the smooth sine-type $I_{tun}(\Phi)$ oscillations with significantly reduced magnitude and period $\Delta\Phi/\Phi_0 = 1$ on the same structures, which at larger cross sections $\sigma$ demonstrated qualitatively different oscillation patter, supports the particular size dependent scenario – quantum fluctuations. Here we would like to emphasize once again that the concern about sample homogeneity with respect to interpretation of our data based on QPS scenario is superfluous, how it has been explicitly shown in the recent theoretical analysis[32].

To summarize, we demonstrated that in superconducting aluminum and titanium loop-shaped tunnel nanostructures the period, the magnitude and the shape of tunnel current oscillations are dramatically modified below a certain cross section of the wire forming the loop. The comprehensive evaluation of the data leads to the conclusion that a simple interpretation based on sample inhomogeneity is not supported by microscopic and elemental analysis of the samples, and, what is even more important, has a deficiency in explaining the gradual reduction of the period and the magnitude of the oscillations with reduction of the characteristic dimension (cross section of the wire forming the loop). Alternative scenarios based on thermal fluctuations and/or unusual size-dependent coupling of the samples to external noisy environment are not credible. In our opinion, the most plausible interpretation of the results is related to the essentially size-dependent nanoscale phenomenon - quantum phase slips, suppressing the persistent currents in the ultra-narrow loops. The interpretation is not improbable as the evidence of QPS has been reported in a number of independent experiments on nanowires of similar dimensions, and there are no obvious reasons why the same mechanism should be absent in nanorings. The effect is of fundamental importance for quantum solid state physics in general, and superconductivity - in particular. To our best knowledge, the discovery is the first experimental observation of the impact of quantum fluctuations on the ground state of a macroscopically coherent system. Additionally to the basic science importance, the emerging new physics is expected to result in intriguing applications: quantum standard of electric current [33] and qbit [34].

**METHODS**

The samples were fabricated using electron beam lithography followed by the ultra-high vacuum deposition of the metals and the lift-off technique. The double-junction geometry was used to eliminate the undesired contribution of quasiparticles and to make the measurements less invasive compared to a single-junction N-I-S configuration [22]. The oval shape of the loop electrode was selected to eliminate parasitic overlapping shadows emerging in two-angle metal evaporation process. In earlier samples the loop contained 'ears' at the poles with tunnel junctions (Fig. 2a), while in later designs the loop overlapped the contacts (Fig. 2b).

Contacts (S1) were made of aluminum oxidized in vacuum forming tunnel barriers (I), for the loop (S2) we utilized either aluminum, or titanium. Both materials have already demonstrated non-vanishing contribution of quantum fluctuations deduced from the broadened R(T) dependencies [9,10,12]. Low energetic $Ar^+$ ion beam milling [27,28] was used to gradually and non-invasively reduce the nanostructure line cross section $\sigma$ enabling experimental study of a size-dependent phenomenon. Utilization of $Ar^+$ ions at acceleration voltages $\leq 1$ keV can be considered as virtually introducing no defects as the ion penetration depth $\sim 2$ nm is comparable to the thickness of the naturally grown oxide. Additionally, the sputtering provides polishing effect eroding small scale imperfections of the lift-off fabricated nanostructures[9,10,12].

The dimensions of the fabricated samples were measured with atomic force (AFM) and scanning electron (SEM) microscopes. However the AFM/SEM study after each sputtering would pose a too high threat of damaging the extremely fragile structures. In practice, the AFM/SEM measurements were taken after the first (high dose) sputtering step and finally at the end of the last measuring session. The sample dimensions at the intermediate steps were defined by interpolation of the known initial and final AFM/SEM data. The method results in high relative errors in determination of the average cross section of the line forming the loop. According to our previous studies of aluminium and titanium nanowires[9,10,12] enabling independent determination of their cross section from the normal state resistance, after multiple sessions of ion beam sputtering the sample diameter can be specified with accuracy better than $\pm 3$ nm. The large error in determination of the diameter of the wire forming the loop in tunnel nanostructures comes mainly from the uncertainty in defining the interface between the metal and the sputtered substrate, and not the actual roughness (variation of the cross section) for a given sample (see Fig. 2d, inset).

High resolution transmission electron microscopy (TEM) analysis revealed the expected non-single-crystalline structure of the samples. The grains were compactly packed with inevitable dislocations and associated stacking faults (Fig. 2c). However, no signature of any sort of inclusions (e.g. foreign material clusters) or distortion due to ion implantation in the ion milled samples was found. Elemental analysis of samples, fabricated using the same vacuum chamber as in present work, revealed the highest concentration of foreign elements inside the metal matrix being associated with 0.3 at. % of oxygen[12]. The high-resolution TEM images of samples from the present paper and wide 2D films and nanowires, fabricated under the same conditions, were indistinguishable between themselves, and what is even more important – indistinguishable before and after the ion milling. The observation supports the statement that the utilized reduction of the cross section of the nanowire forming the loop by low energy ion milling cannot introduce any structural defects[12].

After each step of reducing the cross section, the samples were cooled down to temperatures well below the critical temperature of aluminium and/or titanium using $He^3/He^4$ dilution refrigerator. The measurements were performed inside electromagnetically shielded room using battery powered front-end amplifiers and carefully filtered input/output lines. The actual effective electron temperature $T_e$ was determined by fitting the zero-field $I_{tun}(V, B=0)$ dependence with the familiar text-book expression. Four-stage RF filtering enabled us to keep the increase of the electron temperature $\delta T_e \leq 20$ mK above the base temperature of the refrigerator $T_{bath}$[35]. The tunnel current $I_{tun}$ over the whole structure was measured as function of perpendicular magnetic field intensity at several chosen constant bias voltages. The magnetic field sweeps (covering typically 5 to 10 periods of the oscillations) ranged from few up to ~ 40 minutes. The majority of the experiments were made at temperatures between 50 mK and 150 mK, where neither the sweep rate, nor the temperature dependent effects were observed.

The effective area of the loop electrode calculated from the periodicity was found to vary from a cool down to another one less than 0.5 %, which is of the same order as within a single measurement session. The difference between the calculated effective area and the one measured from SEM/AFM analysis was found to be also about 0.5 %. The error in defining the magnetic field due to sample/coil misalignment was estimated to be about 0.05 %, hence being smaller than the above indicated uncertainties.

# References

[1]. Langer, J. S., Ambegaokar, V. Intrinsic resistive transition in narrow superconducting channels. *Phys. Rev.* **164,** 498-510 (1967).

[2]. McCumber, D. E., Halperin, B. I. Time scale of intrinsic resistive fluctuations in thin superconducting wires. *Phys. Rev. B* **1,** 1054-1070 (1970).

[3]. Lukens, J. E. , Warburton, R. J. , Webb, W. W. Onset of Quantized Thermal Fluctuations in "One-Dimensional" Superconductors. *Phys. Rev. Lett.* **25**, 1180-1184 (1970).

[4]. Newbower, R. S., Beasley, M. R., Tinkham, M. Fluctuation Effects on the Superconducting Transition of Tin Whisker Crystals. *Phys. Rev. B* **5**, 864–868 (1972).

[5]. Zaikin, A. D., Golubev, D. S., van Otterlo, A., Zimanyi, G. T. Quantum Phase Slips and Transport in Ultrathin Superconducting Wires. *Phys. Rev. Lett*. **78**, 1552-1555 (1997).

[6]. Golubev, D. S., Zaikin, A. D. Quantum tunneling of the order parameter in superconducting nanowires. *Phys. Rev. B* **64**, 014504-1 – 014504-14 (2001).

[7]. Giordano, N. Evidence for macroscopic quantum tunneling in one-dimensional superconductors. *Phys. Rev. Lett.* **61,** 2137-2140 (1988).

[8]. Bezryadin, A., Lau, C. N., Tinkham, M. Quantum suppression of superconductivity in ultrathin nanowires. *Nature* **404,** 971-974 (2000).

[9]. Zgirski, M., Riikonen, K.-P., Touboltsev, V., Arutyunov, K. Yu. Size dependent breakdown of superconductivity in ultranarrow nanowires. *Nano Lett.* **5,** 1029-1033 (2005).

[10]. Zgirski, M., Riikonen, K.-P., Touboltsev, V., Arutyunov, K. Yu. Quantum fluctuations in ultranarrow superconducting aluminum nanowires. *Phys. Rev. B* **77**, 054508-1 - 054508-6 (2008).

[11]. Arutyunov, K. Yu., Golubev, D. S., Zaikin, A. D. Superconductivity in one dimension. *Phys. Rep.* **464,** 1-70 (2008).

[12] Lehtinen, J. S., Sajavaara, T., Arutyunov K. Yu. and Vasiliev, A. Evidence of quantum phase slip effect in titanium nanowires, arXiv:1106.3852v1 (2011).

[13]. Matveev, K. A. , Larkin, A. I., Glazman, L. I. Persistent current in superconducting nanorings. *Phys. Rev. Lett.* **89,** 096802-1 - 096802-4 (2002).

[14]. Koshnick, N. C., Bluhm, H., Huber, M. E., Moler, K. A. Fluctuation Superconductivity in Mesoscopic Aluminum Rings, *Science* **318**, 1440 – 1443 (2007).

[15] Bert, J. A., Koshnick, N. C., Bluhm, H., Moler, K. A. Fluxoid fluctuations in mesoscopic superconducting rings, arXiv:1008.4821v1.

[16] Berger, J. Influence of Thermal Fluctuations on Uniform and Nonuniform Superconducting Rings according to the Ginzburg-Landau and the Kramer-Watts-Tobin Models. arXiv:0904.2120v3

[17] Bluhm, H. , Koshnick, N. C., Huber, M. E., Moler, K. A., Magnetic response of mesoscopic superconducting rings with two order parameters, *Phys. Rev. Let*. **97**, 237002-237005 (2006)

[18]. Little, W. A., Parks, R. D. Observation of quantum periodicity in the transition temperature of a superconducting cylinder. *Phys. Rev. Lett.* **9,** 9-12 (1962).

[19]. Pedersen, S., Kofod, G. R., Hollingbery, J. C., Sørensen, C. B., Lindelof, P. E. Dilation of the giant vortex state in a mesoscopic superconducting loop. *Phys. Rev. B* **64,** 104522-1 - 104522-4 (2001).

[20]. Vodolazov, D. Y., Peeters, F. M. , Dubonos, S. V., Geim, A. K. Multiple flux jumps and irreversible behavior of thin Al superconducting rings. *Phys. Rev. B* **67,** 054506-1 - 054506-6 (2003).

[21]. Anthore, A., Pothier, H., Esteve, D. Density of States in a Superconductor Carrying a Supercurrent. *Phys. Rev. Lett*. **90**, 127001-127005 (2003).

[22]. Arutyunov, K. Yu. , Hongisto, T. T. Normal-metal-superconductor interferometer. *Phys. Rev. B* **70**, 064514-1 - 064514-6 (2004).

[23]. Vodolazov, D. Y., Peeters, F. M. , Hongisto, T. T. , Arutyunov, K. Yu. Microscopic model for multiple flux transitions in mesoscopic superconducting loops. *Europhys. Lett.* **75 (2),** 315-320 (2006).

[24] Giazotto, F. , Heikkilä, T. T., Luukanen, A. , Savin, A. M. and Pekola, J. P. Opportunities for mesoscopics in thermometry and refrigeration: Physics and applications, *Rev. Mod. Phys*. **78**, 217-274 (2006)

[25] Tirelli, S., Savin, A. M., Pascual Garcia, C., Pekola, J. P., Beltram, F., Giazotto, F. Manipulation and Generation of Supercurrent in Out-of-Equilibrium Josephson Tunnel Nanojunctions. *Phys. Rev. Lett.* **101**, 077004-1 − 0777004-4 (2008).

[26] P. Li, P. M. Wu, Y. Bomze, I. V. Borzenets, G. Finkelstein, and A. M. Chang, Retrapping current, self-heating, and hysteretic current-voltage characteristics in ultranarrow superconducting aluminum nanowires, *Phys. Rev. B* 84, 184508 (2011).

[27]. Savolainen, M., Touboltsev, V., Koppinen, P., Riikonen, K.-P., Arutyunov, K. Yu. Ion beam sputtering for progressive reduction of nanostructures dimensions. *Appl. Phys. A* **79,** 1769-1773 (2004).

[28] Zgirski, M., Riikonen, K.-P. , Tuboltsev, V. , Jalkanen, P. , Hongisto, T. T. , Arutyunov, K. Yu. Ion beam shaping and downsizing of nanostructures. *Nanotechnology* **19,** 055301-1 - 055301-6 (2008).

[29]. Zgirski, M., Arutyunov, K. Yu. Experimental limits of the observation of thermally activated phase-slip mechanism in superconducting nanowires. *Phys. Rev. B* **75,** 172509-1 - 172509 -4 (2007).

[30]. Oreg, Y., Finkel'stein, A. M. Suppression of $T_c$ in Superconducting Amorphous Wires. *Phys. Rev. Lett.* **83**, 191-194 (1999).

[31]. Zhou, F. , Spivak, B. Superconducting Glass State in Disordered Thin Films in a Parallel Magnetic Field. *Phys. Rev. Lett*. **80**, 5647–5650 (1998).

[32] Vanevic, M., Nazarov, Yu. V. Quantum phase slips in superconducting wires with weak links, arXiv:1108.3553 (2011).

[33]. Mooij, J. E. , Nazarov, Yu. V. Superconducting nanowires as quantum phase-slip junctions. *Nature Physics* **2,** 169 -172 (2006).

[34]. Mooij, J. E., Harmans, C. J. P. M. Phase-slip flux qubits. *New J. Phys.* **7,** 219-1 – 219-7 (2005).

[35] . Arutyunov, K. Yu.,  Auraneva, H. -P. and Vasenko, A. S. Spatially resolved measurements of nonequlibrium quasiparticle relaxation in superconducting aluminium, *Phys. Rev. B* **83**, 104509-1 - 104509-7 (2011).

## Acknowledgements

Authors would like to acknowledge O. Astafiev, F. Hekking, Y. Nazarov, T. Klapwijk, D. Vodolazov, A. Zaikin and A. Zorin for valuable discussions; M. Zgirski and P. Jalkanen for assistance with ion beam treatment. The work was supported by the Finnish Academy project FUNANO, Finnish Technical Academy project DEMAPP and Ministry of Science and Education of Russian Federation grant 2010-1.5-508-005-037.

**Author contributions:** T.T.H. designed and fabricated the structures, performed experiments and analyzed the data.  J.S.L. fabricated structures and participated in measurements. L.I.L. made AFM analysis. A.L.V. contributed with TEM analysis. K.Yu.A. planned the project, made numerical simulations and interpreted the results.

# Figure Legends

Fig. 1. **Schematic dependence of energy $E$, persistent current $I_S$ and tunnel current $I_{tun}$ on normalized magnetic flux $\Phi/\Phi_0$.** (a) Quantized energy spectrum $E_n$ of a thin-walled superconducting ring. Solid thick line corresponds to the evolution of the system following the ground state, dashed line – metastable states, and dots – the limit of quantum fluctuations. (b) Corresponding dependence of the persistent current $I_s$. Horizontal dashed line represents the critical current $I_c$ setting the ultimate limit for the magnitude of a persistent current circulating in the loop. (c) Corresponding dependence of the tunnel current $I_{tun}$.

Fig. 2. **Microscopic images of the structures**. (a) SEM image of a S1-I-S2-I-S1 tunnel structure with the loop having 'ears' at the poles; and schematic of measurements. (b) SEM image of sputtered Al-AlO$_x$-Ti-AlO$_x$-Al structure without 'ears'. The inset shows the magnified view of the equatorial part of the loop. Blurred bright edge is the pedestal formed by the sputtered Si/SiO$_x$ substrate. (c) TEM image of the thinnest part of a typical Ti structure after the cross section reduction using ion milling. The polycrystalline structure of compactly packed grains with the average size ~ 3 nm does not contain obvious defects and is indistinguishable from the original (non ion milled) samples. The bright layer on top of both the Ti and the Si substrate is Pt deposited after the measurements solely for the purposes of the TEM analysis. (d) AFM image of the thinnest (equatorial) part of Ti loop after ion milling. Inset demonstrates the spatial variation of the height measured along the same sample. Note the extreme smoothness of the surface ± 2 nm originating from the polishing effect provided by the low energetic Ar$^+$ ions.

Fig. 3. **Oscillations of the normalized tunnel current $\left| I_{tun}(\Phi) - I_{tun}^{min} \right| / I_{tun}^{min}$ in external magnetic flux $\Phi/\Phi_0$ of three Al-AlO$_x$-Al-AlO$_x$-Al structures with the same area of the aluminum loop $S$=19.6 $\mu$m$^2$.** Experimental data are shown by circles (○), calculations – by lines. (a) Large period ($\Delta\Phi/\Phi_0 = 8$) and magnitude ($\Delta I_{tun}/ I_{tun}^{min} \sim 0.8$ ) oscillations in the structure with loop formed by 110 nm × 75 nm wire, $V_{bias}$=780 $\mu$V, $T_{bath}$=65±5 mK, $T_e$=70 mK, $\sigma_{fit}^{1/2}$=90.8 nm. The monotonous increase of the base line is due to the gap reduction by magnetic field noticeable at biases close to the gap edge $eV_{bias}\sim2(\Delta_1+\Delta_2)$. (b) Oscillations with variable period in the narrower (ion milled) sample, $T_{bath}$=52 ±5mK, $V_{bias}$=608 $\mu$V, $\sigma^{1/2}$=42±30 nm. Solid line represents calculations at the intermediate limit with T$_e$=70 mK and $\sigma_{fit}^{1/2}$=12.49 nm resulting in $\Delta\Phi/\Phi_0$=3 period, dashed line - calculated $\Delta\Phi/\Phi_0$=1 oscillations in a slightly narrower loop $\sigma_{fit}^{1/2}$=12.37 nm. (c) Same sample as in (b) further gently ion milled at $T_{bath}$=54±5 mK and $V_{bias}$=666 μV. Solid line corresponds to calculations in the QPS limit with $\sigma_{fit}^{1/2}$=12.15 nm and $T_e$=70 mK with the same parameters used to fit $R(T)$ dependencies of Al nanowires [9,10]. Left insets - flux dependencies of the characteristic energies: superconducting pairing potential $\Delta$ (□), spectral gap $G$ (◇), deparing energy $\Gamma$ (▲), rate of quantum fluctuations $E_{QPS}$ (●) and the corresponding thermal energy $k_BT$ (-). Right insets – calculated flux dependence of the persistent current $I_s(\Phi)$ normalized by the critical current $I_c(0)$. Note the change of the shape, reduction of the magnitude and the period of both the tunnel and the persistent current oscillations. In the classic (a) and intermediate (b) regimes the periodicity of oscillations is defined by Eq. (1), while in the essentially QPS regime (c) – by Eq. (4). For details – see the text.

Fig. 4. **Tunnel current $I_{tun}$ oscillations in external magnetic flux $\Phi/\Phi_0$ of Al-AlO$_x$-Ti-AlO$_x$-Al structure with the area of the titanium loop $S$=18.9 μm$^2$**. Top data - for non ion milled sample $\sigma^{1/2}$=66±3 nm, bottom data - for sputtered sample $\sigma^{1/2}$=62±8 nm, $T_{bath}$=65±5 mK at the same bias $V_{bias}$=105 μV. Solid lines represent QPS limit calculations assuming T$_e$=70 mK, $\sigma_{fit}^{1/2}$=63 and 56 nm, respectively, and the same parameters used to fit $R(T)$ dependencies of Ti nanowires[12]. Note the drop of the oscillation amplitude by a factor of ~6 when the diameter $\sigma^{1/2}$ is reduced just by ~ 10%. Inset shows simulation: for the same fitting parameters of the ion milled sample - what would be the tunnel current oscillations in the intermediate limit (similar to Fig. 3b). For details – see the text.

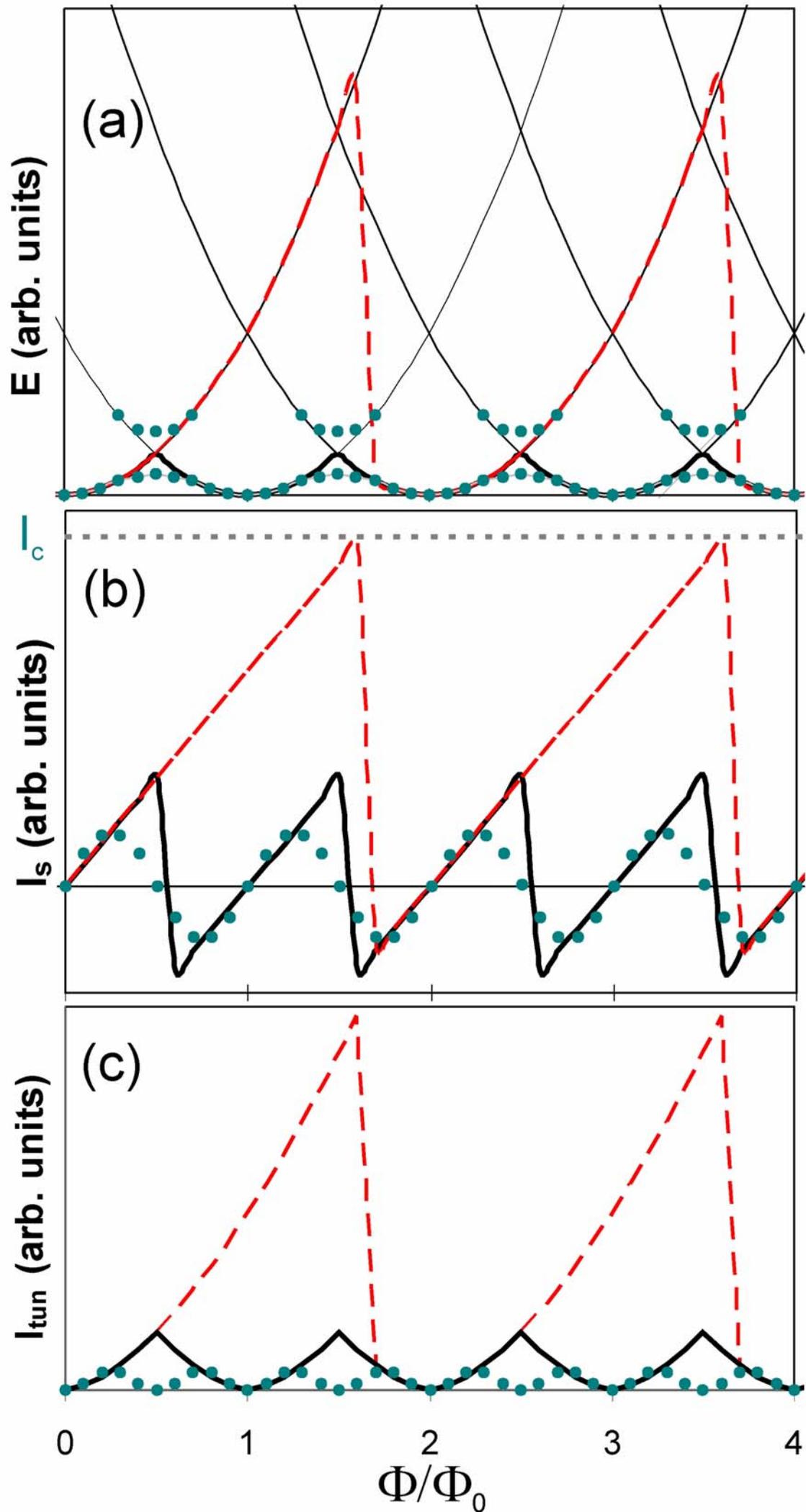

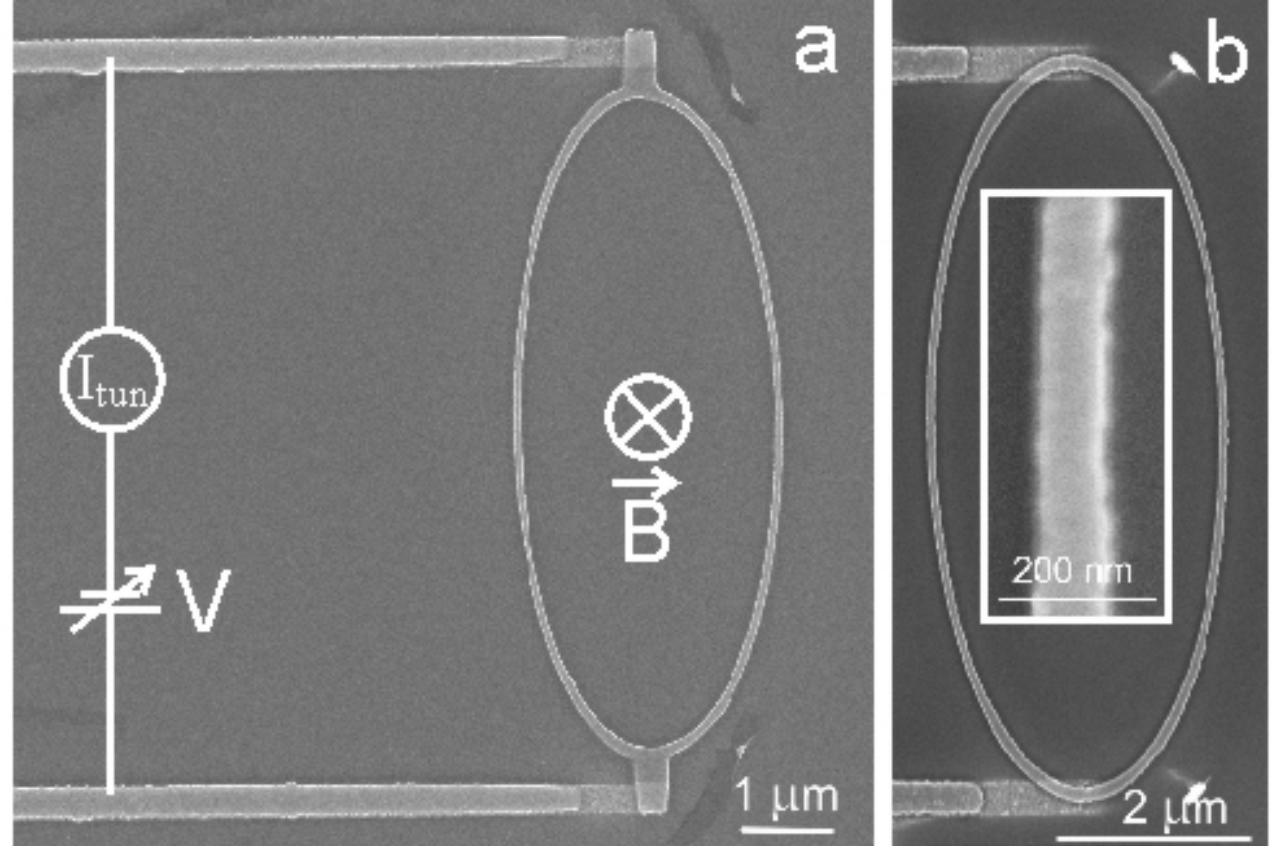

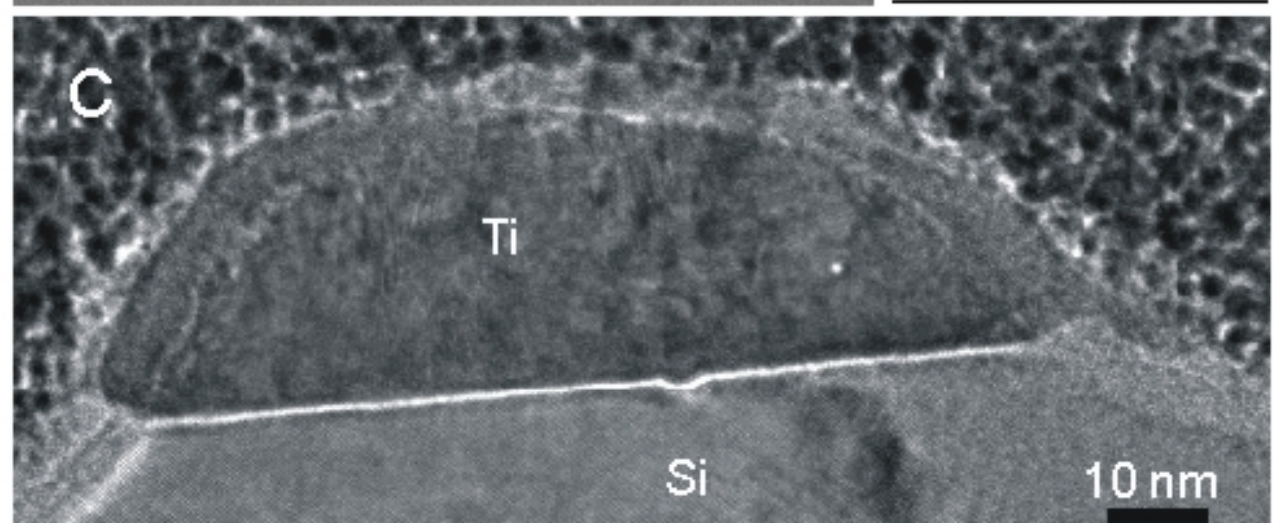

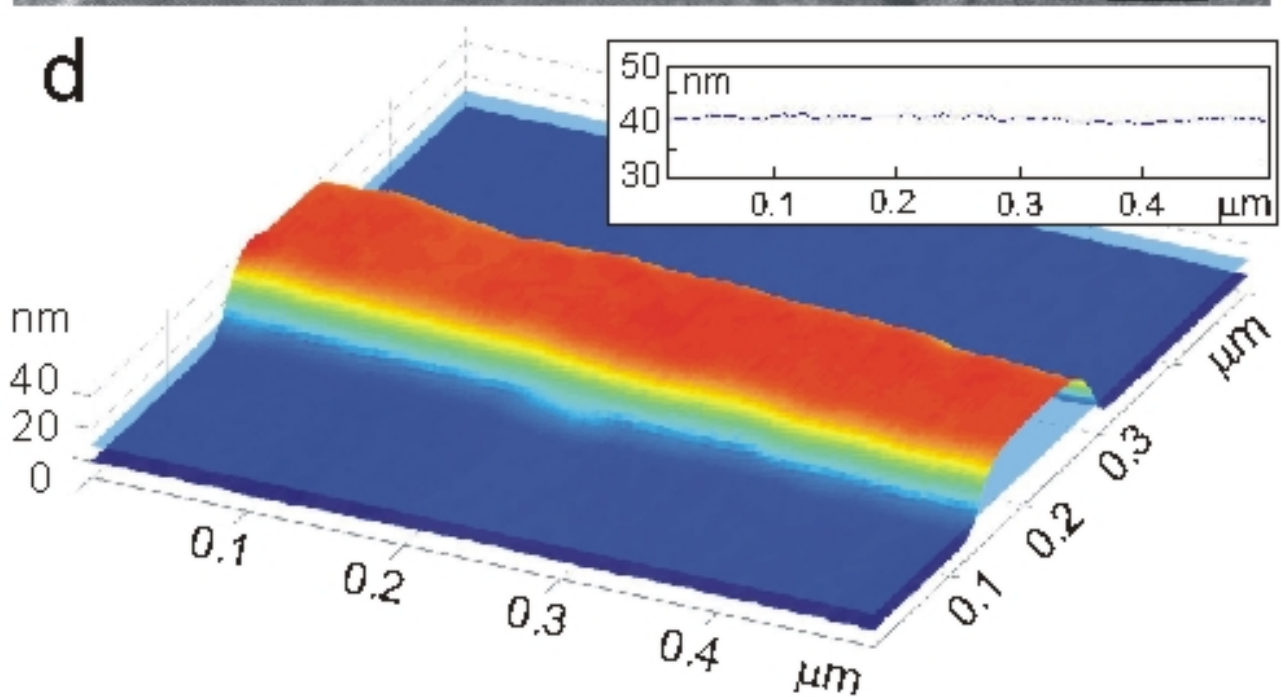

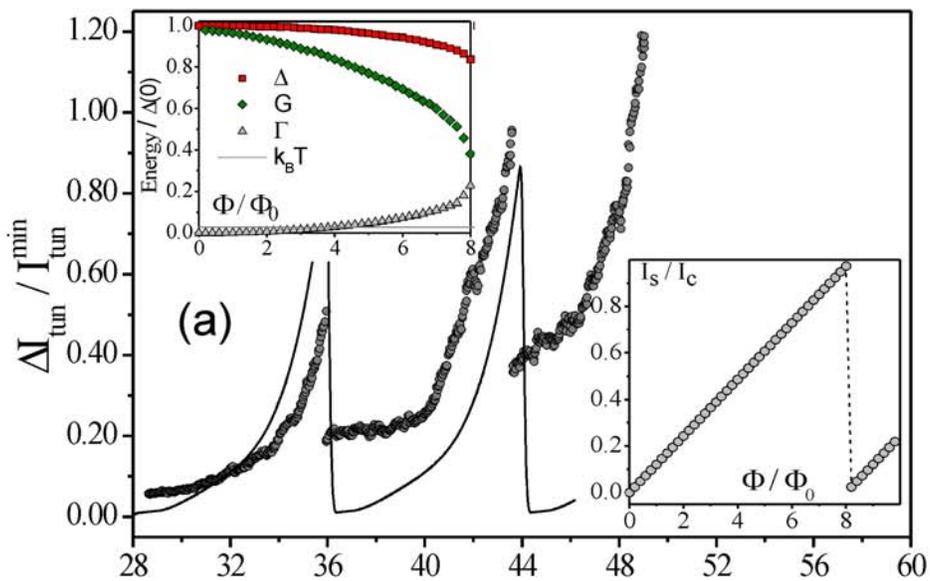

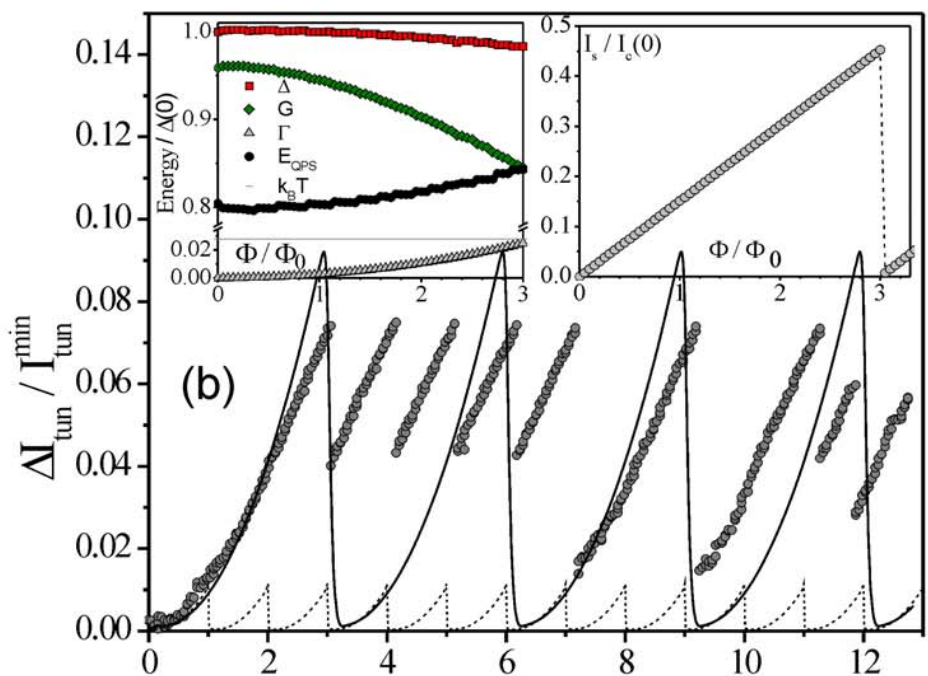

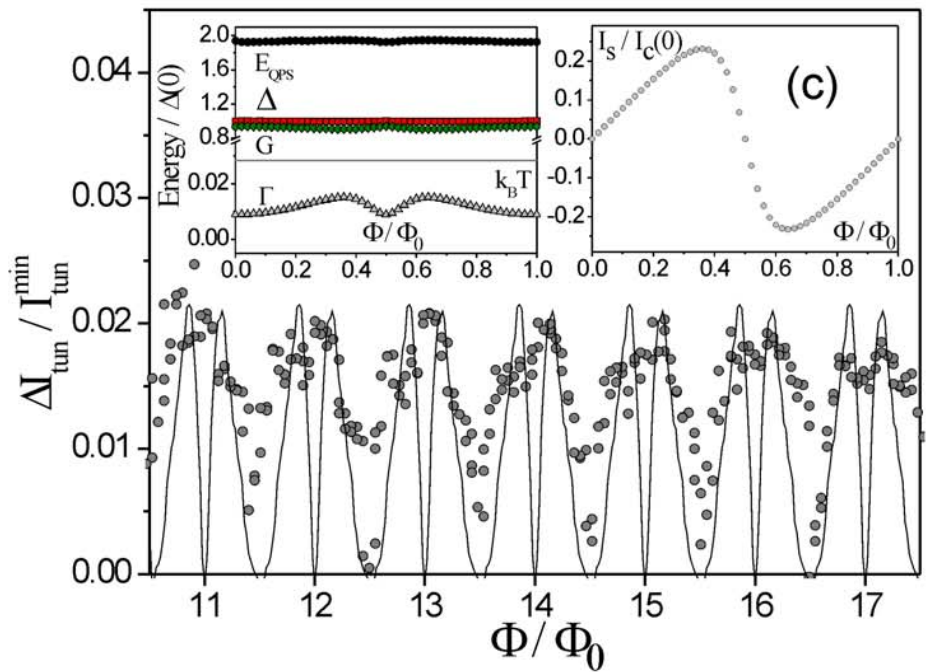

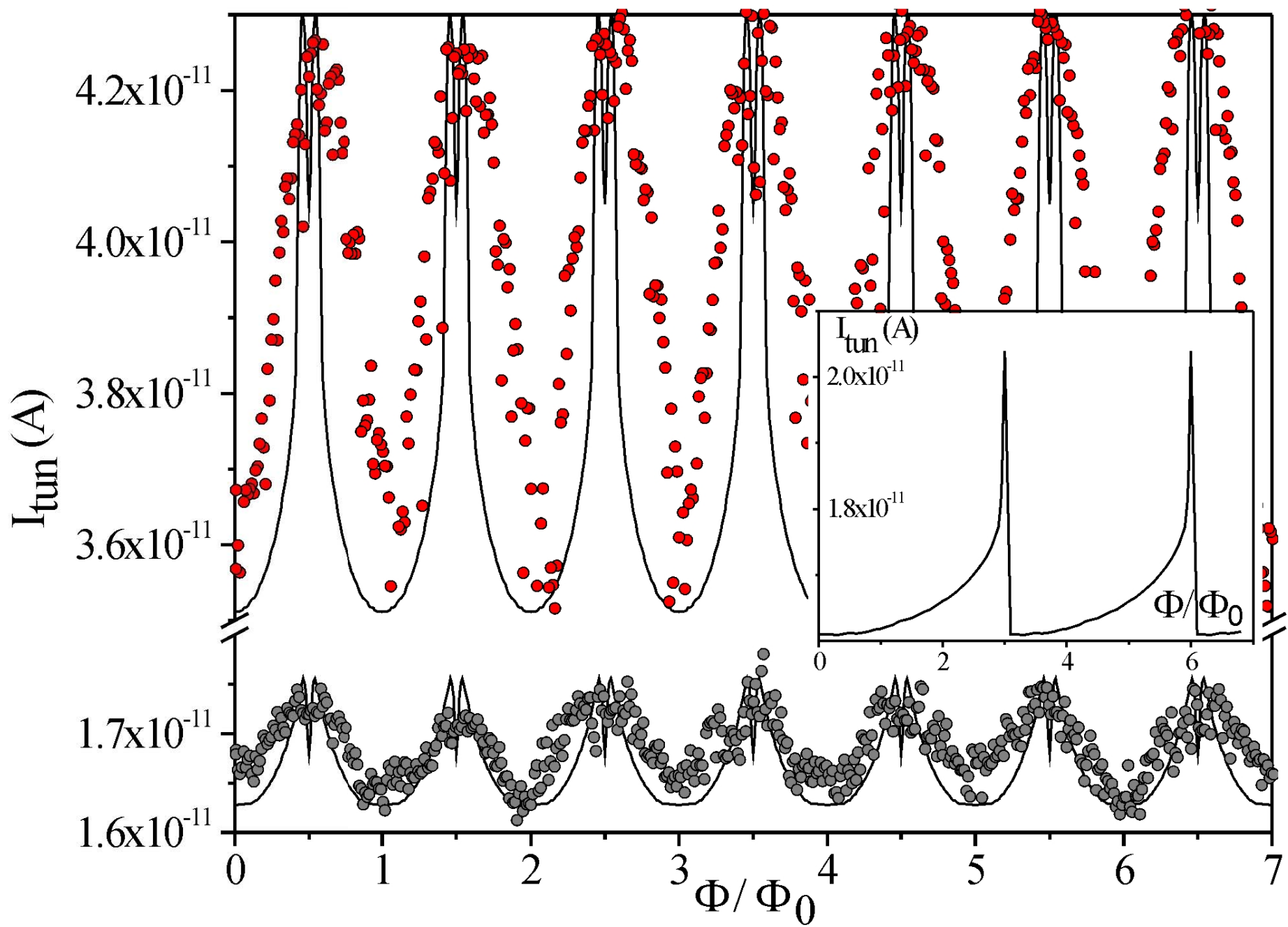